\documentstyle[12pt,epsf]{article}
\newcommand{\be}{\begin{equation}}
\newcommand{\ee}{\end{equation}}
\newcommand{\ba}{\begin{eqnarray}}
\newcommand{\ea}{\end{eqnarray}}
\newcommand{\ft}{\tilde F_{\mu}}

\newcommand{\fmn}{F_{\mu\nu}}
\newcommand{\oh}{\displaystyle{\frac{1}{2}}}
\begin{document}
\title     {Virtual photon effects
                 in $D=3$ Born-Infeld theory.}
\author{C.D.\ Fosco$^a$\thanks{CONICET}\\
and\\
F.A.\ Schaposnik$^b$\thanks{Investigador CICBA, Argentina}
\\
{\normalsize\it
$^a$Centro At\'omico Bariloche,
8400 Bariloche, Argentina}\\
{\normalsize\it
$^b$Departamento de F\'\i sica, Universidad Nacional de La Plata}\\
{\normalsize\it
C.C. 67, 1900 La Plata, Argentina}}
\date{\hfill}
\maketitle
\begin{abstract}
One loop effects due to virtual gauge field propagation in $2+1$
dimensional Born-Infeld theory are investigated. Although this
field theory model is not power counting renormalizable, it can be
consistently interpreted as an effective field theory. We derive the
one-loop effective action in this framework.

Halpern's field strength's formulation is then applied to derive an
effective description for the interaction between magnetically charged
particles, when the gauge field dynamics is determined by a Born-Infeld
action. We compare the results with those of  the 
Maxwell theory.
\end{abstract}

\bigskip

\newpage
\section{Introduction}
$QED_3$-like theories, consisting of charged matter fields in
interaction
with Abelian gauge fields in $2+1$  dimensions, have multiple
applications in both Condensed Matter~\cite{frad} and High Energy
Physics~\cite{jack}.
Even after imposing the requisites of gauge invariance and
renormalizability, the gauge field action is not uniquely determined,
because in an odd number of space-time dimensions it is always possible
to
combine the standard Maxwell Lagrangian with a Chern-Simons term. This
gives rise to interesting, distinctive phenomena.

A different, non quadratic Born-Infeld action
\cite{B}-\cite{b} for
the gauge field in $2+1$ dimensions has been 
recently considered~\cite{more}, in order to
study
the effects its presence has on the properties of 
classical vortex configurations in
a spontaneously broken gauge theory. Now,
the Born-Infeld action induces
non-renormalizable interactions so
that a question that immediately arises  is
  how to deal with the gauge field self-interactions,
at the {\em quantum\/} level. The relevance of these self-interactions
is already evident at the classical level: they provide a mechanism to
avoid the divergence of the electric field produced by a point-like
electric source, and it is certainly worth investigating what could
these
effects be at the quantum level. Namely, we would like to know, for
example, to what extent the field distribution around a point-like
charge
is altered by quantum effects.

The study of loop effects due to non-linearities in a gauge field
action is not a new subject. Indeed, in a precursor work
by Halter~\cite{halt}, a photon loop in the Euler-Heisenberg
Lagrangian~\cite{eule} for $QED_4$ was calculated in an effective field
theory framework. That the theory is an 
{\em effective\/} theory is an important fact since,
being quartic in the field strength,
the
Euler-Heisenberg Lagrangian has coupling constants
with negative mass dimensions and hence is not renormalizable. 
There is however an intrinsic momentum cutoff, given by the mass
of the charged particle that has been integrated out.
The situation is somewhat similar to that taking
place in chiral perturbation
theory, where a non-renormalizable non-linear $\sigma$-model
is used to calculate loops,   yielding relevant information and
having predictive power if only a few terms in the expansion are
calculated.

In this paper, we shall consider the Born-Infeld action
in $2+1$ dimensions (i.e. just the pure
gauge sector of the model discussed in
reference~\cite{more}) as  a `classical' effective theory for the
gauge field $A_\mu$, valid only below a certain
momentum cutoff. This is consistent with the usual constraint in
Born-Infeld theory, which states that the field strength 
should be smaller
than a maximum value  called the ``absolute field''.
Indeed, following a suggestion by Born, there have been
attempts~\cite{hagi} to interpret the Born-Infeld theory as arising
from integrating out some unknown massive matter field, keeping only
the terms containing $F_{\mu\nu}$, and no extra derivatives.
We shall, in this paper, adopt this point of view, namely,
to treat this kind of Lagrangian as if it came from a low-momentum
expansion. It will then be reliable only up to a certain momentum
cutoff,
of the order of the mass of the original
fields that have been integrated out.

We may call the Born-Infeld theory for $A_\mu$ ``classical''
since $A_\mu$ has not yet been quantized. The next step in an effective
low momentum description amounts to adding  to this  ``classical''
effective theory  the quantum  effects due to  virtual
photons~\cite{halt},
calculating  the  corresponding  one-loop ``quantum effective action''.
We will compute this one loop correction neglecting derivatives
of $F_{\mu \nu}$, since they have already been disregarded at the
`classical' level, to avoid higher derivatives and non-causal
propagation. This amounts to evaluating the effective one-loop
action in the presence of a constant $\fmn$. This kind
of calculation, for the case of a general quartic gauge
field action in $2+1$ dimensions has been presented in
\cite{aitc}.

A different approach is followed in the second part of this
paper We consider there the Born-Infeld theory in the so-called
field-strength formalism. This formulation, proposed by Halpern
in the seventies~\cite{hal}, amounts to a kind
of dual first-order formulation. We shall use it in order to derive
a `dual' Born-Infeld action, appropriate to the description of
the interaction between magnetically charged particles.


This paper  is organized  as follows:  In section  2 the
one-loop quantum correction to the classical Born-Infeld action
for the gauge field is defined and calculated. In section 3 we
derive an effective theory for the interaction between
magnetically charged particles.

\section{One-loop effective action}
As we shall study here the quantum effects due to gauge
field fluctuations, we  consider the generating functional
\be
{\cal Z}[J_\mu] \;=\; \int \,{\cal D} A_\mu \, \exp\left(
{i S_{BI} [A] - i
\int d^3 x J_\mu (x) A^\mu (x) }
\right)
\label{dz}
\ee
where
\be
S_{BI} [A] \;=\; - \beta^2 \, \int d^3 x \left[
R (\frac{F^2}{2 \beta^2})\,-\,1 \right]
\label{sbi}
\ee
Here $R(x)= \sqrt{ 1 + x}$,  $F^2 \,\equiv\,
F_{\mu\nu}F^{\mu\nu}$, $\beta$ is a constant with mass dimensions
$[\beta] = m ^{3/2}$. It should be noted that the constraint
\be
F^2 < 2 \beta^2
\label{const}
\ee
assures the reality of the classical action, a sensible
physical constraint, if we want to have unitary time
evolution. The constant $\beta$ shall be used to fix the
momentum cutoff, to a value $\beta^{{2}/{3}}$.
Of course, the momentum cutoff could be not exactly equal to
$\beta$, but must be of the same order~\footnote{Note
that $\beta$ is the only dimensionful constant we have.}.

For $F^2 < 2 \beta^2$, an expansion of (\ref{const}) yields
\be
S_{BI} [A] \;=\; - \, \int d^3 x \left[
\frac{1}{4}F^2 - \frac{1}{32 \beta^2} (F^2)^2
+ \frac{1}{128 \beta^4} (F^2)^3 + {\cal O}(\,
\frac{(F^2)^4}{\beta^6}\,)\right]
\ee
where the fact that vertices with an arbitrary large number of
derivatives arise is evident. Those vertices are responsible for
the non-renormalizability.

To calculate the one-loop effective action for this model,
we rotate to Euclidean space. The Euclidean
version of (\ref{dz}) takes then the form
\be
{\cal Z}[J_\mu] \;=\; \int \,{\cal D} A_\mu \, \exp \left(
{- S [A] +
\int d^3 x J_\mu (x) A_\mu (x) }
\right)
\ee
where
\be
S [A] \;=\; + \beta^2 \, \int d^3 x \left[
R(\frac{F^2}{2 \beta^2}) \,-\,1 \right]
\label{secl}
\ee
where now $F^2 \equiv F_{\mu\nu}F_{\mu\nu}$ is positive definite.
The Euclidean effective action to one-loop order is, as usual, given
by the following expression
\be
\Gamma_{eff}[A] \;=\; S[A] \;+\; \Gamma^{(1)}[A]
\label{geff}
\ee
where $S[A]$ is the classical Euclidean action (\ref{secl}), and
\be
\Gamma^{(1)}[A] \;=\; \oh \left. {\rm Tr}
\ln (\frac{\delta^2 S}{\delta A_\mu
\delta A_\nu}) \,-\, \oh {\rm Tr} \ln
(\frac{\delta^2 S}{\delta A_\mu \delta A_\nu})\right|_{A = 0}
\;.
\label{g1}
\ee
In computing the second functional derivative of $S$, we drop terms
containing derivatives of $\fmn$ thus obtaining
$$\frac{\delta^2 S}{\delta 
A_\mu (v) \delta A_\nu (w)} |_{\fmn = const.}
= $$
\be 2 \, \left\{ \, R'(\frac{F^2}{\beta^2})\,
(-\partial^2 \delta_{\mu \nu}+\partial_\mu \partial_\nu)
 - \frac{1}{\alpha} \partial_\mu \partial_\nu
-\frac{2}{\beta^2} \, R''(\frac{F^2}{\beta^2})\,
F_{\mu \alpha} F_{\nu \beta}\, \partial_\alpha \partial_\beta \right\}
\delta
(v - w) \,,
\label{sfd}
\ee
where $v ,\, w$ are 
coordinates of two space-time points .

$\Gamma^{(1)} = \Gamma^{(1)}[F] $ denotes 
the part  of the total effective action that
includes one loop quantum effects coming from $A_\mu$. 
$$ \Gamma^{(1)} [F] \;=\; \oh \, {\rm Tr} \, \ln
\left\{
R'(\frac{F^2}{2\beta^2})\;
(-\partial^2 \delta_{\mu \nu}+\partial_\mu \partial_\nu)
 - \frac{1}{\alpha} \partial_\mu \partial_\nu - \right.
$$
\be
\left.\frac{2}{\beta^2} \; R''(\frac{F^2}{2\beta^2})\;
F_{\mu \alpha} \; F_{\nu \beta}\partial_\alpha \partial_\beta
\right\}
-\oh {\rm Tr} \ln \left\{
(-\partial^2 \delta_{\mu\nu} +\partial_\mu \partial_\nu) -
\frac{1}{\alpha}
\partial_\mu \partial_\nu
\right\} \;,
\label{v1}
\ee
where the usual field independent infinite constant has been
subtracted.
The symbol `${\rm Tr}$' in (\ref{v1}) means 
both functional and discrete
trace (over the indices $\mu$, $\nu$), and care should be taken of
the fact that the log must be understood as the logarithm of a
matrix function. Passing to momentum space, we obtain
$$
\Gamma^{(1)}[F]\;=\; \int d^3 x \int \frac{d^3 k}{(2 \pi)^3} \,
\left\{ {\rm tr}
[ R' (k^2 \delta_{\mu\nu} - k_\mu k_\nu) \right.
$$
\be
\left.+ \frac{1}{\alpha} k_\mu k_\nu + \frac{2 R''}{\beta^2}
F_{\mu\alpha} F_{\nu\beta} k_\alpha k_\beta]
 -{\rm tr} [ (k^2 \delta_{\mu\nu} - k_\mu k_\nu)
+ \frac{1}{\alpha} k_\mu k_\nu ]
\right\} \;.
\label{ge1}
\ee
To take the trace over the discrete indices we just calculate
the eigenvalues of the corresponding three by three matrix.
After doing that,  and cancelling like terms between the two
lines of (\ref{ge1}), we get
\be
\Gamma^{(1)}[F]\;=\; \oh \int d^3 x \int \frac{d^3 k}{(2 \pi)^3} \,
\left\{ 2 \ln R'
\,+\, \ln[1 + \frac{2 R''}{\beta^2 R'}
F_{\mu\alpha} F_{\mu\beta} \frac{k_\alpha k_\beta}{k^2}]
\right\} \;,
\ee
where we must keep in mind that the momenta must be integrated
inside a ball of radius $\beta^{\frac{2}{3}}$, the momentum
cutoff.
This integration over momentum yields, after some straightforward
algebra,
\be
\Gamma^{(1)}[F] \;=\; - \frac{\beta^2}{6 \pi^2}
\int d^3 x \ln \sqrt{1 + \frac{F^2}{2 \beta^2}}
\,+\, \frac{\beta^2}{6 \pi^2} \int d^3 x
\left[
\frac{{\rm arctan}(\sqrt{\frac{F^2}{2 \beta^2}})}{
\sqrt{\frac{F^2}{2 \beta^2}}}
-1 \right] \;.
\ee

Rotating back to Minkowski space-time, we obtain for the
full one loop effective action
$$
\Gamma_{eff}[F] \;=\; - \beta^2 \, \int d^3 x \left[
R (\frac{F^2}{2 \beta^2}) \,-\,1 \right]
$$
\be
+ \frac{\beta^2}{6 \pi^2} \int d^3 x \ln
\sqrt{1 + \frac{F^2}{2 \beta^2}}
\,-\, \frac{\beta^2}{6 \pi^2} \int d^3 x
\left[
\frac{{\rm arctan}
(\sqrt{\frac{F^2}{2 \beta^2}})}{\sqrt{\frac{F^2}{2 \beta^2}}}
-1 \right] \;,
\label{sem}
\ee
or, in a more compact form,
\be
\Gamma_{eff}[F] \;=\; - \beta^2 \, \int d^3 x \left[
R_{eff} (\frac{F^2}{2 \beta^2}) \,-\,1 \right]
\ee
where
\be
R_{eff} (x) \;=\; R(x) - \frac{1}{6 \pi^2} \ln \sqrt{1 + x}
\,+\, \frac{1}{6 \pi^2}
\left[\frac{{\rm arctan}(\sqrt{x})}{\sqrt{x}}-1 \right] \;,
\ee

Let us now discuss the behaviour of the extra terms in (\ref{sem})
that modify the Born-Infeld action. A large $\beta$ expansion
of the classical Born-Infeld action yields the Maxwell action,
while for the quantum corrected version (\ref{sem}) we obtain
\be
\Gamma_{eff}[F] \;=\; \int d^3 x \left[ - (\frac{1}{4} -
\frac{5}{72 \pi^2}) F^2 \,+\,
{\cal O} ( (\frac{F^2}{\beta^2})^2 ) \right],
\label{inde}
\ee
a very small modification indeed. In the opposite regime,
we have instead
\be
\Gamma_{eff}[F] \;=\; \beta^2 \int d^3 x \, \left[-\sqrt{\frac{F^2}{2
\beta^2}} + {\frac{\ln ({\frac{\sqrt{F^2}}
{2 \beta^2}})}{6\,{{\pi }^2}}}
+ (1 + {\frac{1}{6\,{{\pi }^2}}}) \,+
\,{\cal O}(\sqrt{\frac{\beta^2}{F^2}}) \right]
\;,
\ee
where we see that the corrections grow at most logarithmically
with $|F|$, while the classical action goes like $|F|$.

It is interesting to note at this point that Born original idea
was not only to have a classical theory for charged particles
without the infinite self-energy problem of Maxwell theory but also
to compare the quantum answer of this last theory (i.e. the Euler Heisenberg
effective action) with the classical ones arising from his non-linear
electromagnetism which could account for the mass of the electron
in terms of its electromagnetic energy. In this respect, Born \cite{b}
disregarded the discrepancy between the factors multiplying
the quartic terms in the expansion of his non-linear classical
Lagrangian and that of Euler-Heisenberg and just adjusted $\beta$
in order to have an overall agreement. In
this context, including one loop-effects
in the Born-Infeld theory   incorporates
some quantum effects that should be comparable with those
arising from a similar calculation in the Euler-Heisenberg
effective theory (as done by Halter in the $d=4$ case \cite{halt}).

The effective Lagrangian in Minkowski space is (see eq.(15))
\be
L_{eff} \;=\;   L_{BI} + L_{q}
\;=\; - \beta^2  \left[
R_{eff} (\frac{F^2}{2 \beta^2}) \,-\,1 \right]
\label{a1}
\ee
with $R_{eff}$ given by eq.(16)
\be
R_{eff} (x) \;=\; \sqrt{1 + x}
  - \frac{1}{6 \pi^2} \ln \sqrt{1 + x}
\,+\, \frac{1}{6 \pi^2}
\left[\frac{{\rm arctan}(\sqrt{x})}{\sqrt{x}}-1 \right]
\label{a2}
\ee
As usual, we define
\be
E^i = F^{0i} \;\;\;\; B = \varepsilon_{ij}F^{ij}
\label{a3}
\ee
so that
\be
x = \frac{F^2}{2 \beta^2} = \frac{1}{\beta^2}(B^2 - E^2)
\label{a4}
\ee
with $E^2 = E^iE^i$.

It is interesting at this point to
 define  electric and magnetic polarization vectors for the
vacuum as
\be
P_i \equiv \frac{\partial L_{q} }{\partial E^i}
\label{a5}
\ee
\be
M \equiv \frac{\partial L_q }{\partial B}
\label{a6}
\ee
We obtain from (\ref{a1}) 
\be
P^i = \chi E^i \;\;\;\; M = -\chi B
\label{a7}
\ee
with the susceptibility $\chi$ taking the form
\be
\chi(x) = \frac{1}{6\pi^2}\left[ \frac{1-x}{x(1+x)}
-\frac{\arctan (\sqrt{x})}{x^{3/2}}
\right]
\label{a8}
\ee
We thus see that the vacuum supports magnetic and electric
polarization.  As it also happens for the Euler-Heisenberg 
effective theory \cite{L}, these
polarization vectors  vanish whenever $E^2 = B^2$ since $\chi(0) = 0$.
Moreover, this is in agreement with the conditions imposed long ago
by Weisskopf \cite{wei} on the vacuum polarization arising
in the study of Euler-Heisenberg effective actions, namely that
the energy density and the susceptibility
have to vanish in the absence of fields in
order to have a consistent effective theory.

Let us now analyse   the electric field
distribution around a point-like charge 
when one takes into account the contribution of
$\Gamma^{(1)}$ to the Born-Infeld effective
action. The Gauss law for the classical
$2+1$ Born-Infeld theory, for the case of a delta-like distribution of
charge $q$, located at the origin yields,
\be
\partial^j [\frac{E_j}{R(-\frac{E^2}{\beta^2})}] = q \delta (x) \;.
\ee
It is trivial to solve this equation, this leading to
 a solution corresponding to a radial electric
field $E (r)$  of the form
\be
\frac{E(r)}{\sqrt{1-\frac{E^2(r)}{\beta^2}}}\;=\;\frac{q}{2 \pi r} \;.
\ee

For the quantum corrected version the equivalent result is
\be
2 E(r) R'_{eff}(-\frac{E^2(r)}{\beta^2}) \;=\; \frac{q}{2 \pi r} \;,
\label{dist1}
\ee
where
\be
R'_{eff} (x) \;=\;{\frac{1}
     {2\,{\sqrt{1 + x}}}} -
  \frac{1}{12\pi^2}\left(
 \frac{1}{ 1 + x } -
 \frac{1}{x\, ( 1 + x) } +
 \frac{\arctan ({\sqrt{x}})}
         {x^{3/2}} \right) 
\label{dist2}
\ee
and $x$ has to be replaced by $-{E^2}/{\beta^2}$.
In order to plot this electric field as a function of $r$, it
is convenient to measure it in units of $\beta$, namely, we define
$u(r) \equiv {E(r)}/{\beta}$. Hence, (\ref{dist1}) implies
\be
u(r) \left\{ \frac{1}{\sqrt{1-u^2(r)}} - \frac{1}{6\pi^2}
[ \frac{1 + u^2(r)}{u^2 (r) (1 - u^2 (r))} -
\frac{{\rm arctanh(|u(r)|)}}{|u(r)|^3}] \right\} =
\frac{q}{2 \pi \beta r}
\label{dist3}
\ee
We note that in the small-$u$ regime, (\ref{dist3}) reduces to
\be
u(r) [ 1 - \frac{5}{18\pi^2} + {\cal O}(u^2)] \;=\;
\frac{q}{2 \pi \beta r}
\ee
namely,so that it reduces to the usual behaviour, except for a
constant renormalization of the charge. This renormalization
constant could actually be read from (\ref{inde}). One can see
in  figure 1 that the electric fields for the point charge
in the classical and quantum corrected versions of the BI
theory qualitatively coincide.

Let us end this section by consider a formulation than
allows to
capture at least part of the non-perturbative dynamics of
Born-Infeld theory. To this end,
 we shall study here a simplified situation.
which corresponds to ignoring the magnetic field in (\ref{sbi}),
what yields a much simpler action $S_{el}$,
\be
S_{el} [A] \;=\; - \beta^2 \, \int d^3 x \left[
R (-\frac{{\vec E}^2}{\beta^2})\,-\,1 \right]  \;.
\label{sel}
\ee
It is already evident that this action decomposes into an integral
of independent actions, one for each spatial point (there
is no spatial propagation):
\be
S_{el} [A] \;=\; m^2 \int d^2 x S_x [A]\;\;,\;\;
S_x [A] \;=\; - m \int_{t_1}^{t_2} dt \, [
\sqrt{1 - \frac{({\vec E}(x,t))^2}{m^3}} -1 ]
\label{dec}
\ee
where we defined $m = \beta^{\frac{2}{3}}$, which has the units of a
mass. A factor of $m^2$ was extracted in order to keep both
$S_{el}$ and $S_x$ explicitly dimensionless, what simplifies the
treatment.

We see from (\ref{dec})  that it is only necessary to understand
$S_x$, regarding $x$ as an index that labels independent fields.
Now, if we use the timelike gauge $A_0 = 0$, this action becomes
\be
S_x [A] \;=\; - m \int_{t_1}^{t_2} dt \, [
\sqrt{1 - \frac{1}{m^3}
(\frac{\partial}{\partial t}{\vec A}(x,t))^2} - 1 ] \;.
\label{sx}
\ee
We realise that this action is equivalent to the one of a relativistic
particle of mass $m$, with "spacetime coordinates"
\be
X^{\mu} (x,t) \;=\;
(t, \frac{1}{m^{\frac{3}{2}}} {\vec A}(x,t)) \;,
\ee
where the factor of $m$ was introduced in order to have the
same dimensionality for all the components (${\rm mass}^{-1}$).
The action $S_x$ is written in a particular parametrization
of the trajectory, that follows from using the time as a parameter
labeling points along the path. A reparametrization invariant form
can be obtained if one introduces a parameter $\tau$,
$t = t(\tau)$, such that the endpoints of the time interval are
unchanged. Then $X^0 = X^0 (\tau)$, and
\be
S_x [A] \;=\;  m (t_2 - t_1) - m
\int_{t_1}^{t_2} d\tau \,\sqrt{ {\dot X}_\mu (x,\tau)
{\dot X}^\mu (x,\tau) } \;,
\ee
where contractions are performed with the usual Minkowskian metric in
the space of the $X^\mu$.
(This action may also be written in quadratic form at the expense
of introducing an einbein~\cite{cohe}).
Using the known results about the quantum relativistic
particle~\cite{cohe}, we can, for example, calculate the amplitude
corresponding to a transition from a given configuration $X_1$ to
another
one $X_2$. Note that 
if the space coordinates $x$ are different, that will
introduce a $\delta$-function. In terms of $A$, we can express the
amplitude for going from a configuration ${\vec A}_1(x_1)$ at a time
$t_1$ to another one ${\vec A}_2(x_2)$ at a time $t_2$ as follows
\begin{eqnarray}
& & \langle {\vec A}_2(x_2,t_2) | {\vec A}_1 (x_1,t_1) \rangle =
\delta (x_2 - x_1) \, \int \frac{d^3 k}{(2 \pi)^3} \,
 \frac{1}{k^2 - m^2 + i \epsilon} \times
\nonumber \\
& & 
\exp\left({-i k^0 (t_2 - t_1) + i \beta^{-3/2} {\vec k} \cdot ({\vec A}_2 (x_2)
-
{\vec A}_1 (x_1))} \right)  
\label{dific}
\end{eqnarray}
Using this technique, many other quanbtities can be calculated 
exactly (like for example the free energy at finite temperature).

It should be whorthwile at this point
to comment on an aspect  of  renormalizability
which has interesting implications. We have
justified the use of a momentum cut-off in order to make sense
from divergent quantities by considering the BI lagrangian as an
effective one arising in the low-energy regime of some unknown theory
In fact, the renewed  interest on BI theory stems from the fact
 it can be
seen as arising as a part of an effective action derived from string theory
\cite{Tse}-\cite{G}. 

Now, as already stressed in \cite{hagi},
the BI Lagrangian can be seen as a possible Euler-Heisenberg effective
Lagrangian derived from a certain renormalizable supersymmetric Lagrangian.
This, together with the original observation of Deser and Pusalowski
\cite{DP} concerning the fact that the BI Lagrangian is among those
non-polynomial lagrangians which admit supersymmetric extensions
suggests that the analysis of supersymmetry in connection with
BI models and renormalizability should be thoroughfully considered.
In the same sense points the remarkable fact that starting
from a SUSY system of minimally coupled spin $1/2$ and spin $0$
particles one achieves agreement between the Euler-Heisenberg 
effective action and the Born-Infeld action up to and including terms
of order $4$ thus resolving the discrepancy, signaled above,
between BI and the effective action for $QED_4$.
We hope to discuss this and other issues related to supersymmetry
and Born-Infeld theory in a forthcoming work.

\section{Field strength formulation and monopoles}
We begin by reviewing Halpern's derivation, with some
differences due to the fact that the action is of the
Born-Infeld type rather than Maxwell.
The generating functional for Euclidean Green's functions of the
Abelian gauge field $A_\mu$, with a classical Euclidean action
$S$ as in (\ref{secl}) is
\be
{\cal Z}[J_\mu] \;=\; \int \,{\cal D} A_\mu \, \exp
\left({- S[A] + \int d^3 x
J_\mu (x) A_\mu (x) }
\right)
\label{defz}
\ee
where $S[A]$ of course satisfies $S[A + \partial \omega]=S[A]$,
for any $\omega$.
We chose to work in terms of $\ft$, the dual of $\fmn$,
defined by $\ft = \epsilon_{\mu\nu\lambda} \partial_\nu A_\lambda$.
Thus
\be
S[A] \;=\; I(\ft)
\label{relsi}
\ee
where $I$ is the functional.
\be
I(\ft) \;=\; \beta^2 \int d^3 x
\left[ R(\frac{\ft^2}{\beta^2}) - 1 \right] \;.
\ee
We now include into (\ref{defz})
the gauge-fixing factor corresponding to the Landau gauge ($\partial
\cdot A = 0$)
\be
{\cal Z}[J_\mu] \;=\; \int \,{\cal D} A_\mu \,
\delta (\partial \cdot A)
\exp \left(
{- S (A) + \int d^3 x J_\mu (x) A_\mu (x) }
\right)
\label{gfixz}
\ee
where we have omitted the field-independent Faddeev-Popov factor
$\det (-\partial^2)$, since in this case it can be absorbed into the
normalization of the integration measure and has no effect on the
Green's functions derived from (\ref{gfixz}).
To obtain a formulation in terms of $\ft$, we introduce 
in (\ref{gfixz}) a `1' written as follows:
\be
1 \;=\; \int \,{\cal D} \ft \, \delta (\ft - \epsilon_{\mu\nu\lambda}
\partial_\nu A_\lambda) \, \delta (\partial \cdot {\tilde F}) \;.
\label{one}
\ee
Note the presence of a delta functional of the Bianchi identity,
which is a consistency condition for the equation
$\ft - \epsilon_{\mu\nu\lambda}\partial_\nu A_\lambda=0$, whose
solutions are relevant to the first delta-function. The meaning
of the inclusion of that factor can be made explicit by means
of the following argument: Consider the rhs of Equation
(\ref{one}), but this time writing both delta-functionals
in terms of functional Fourier transforms:
$$\int \,{\cal D} \ft \, \delta (\ft - \epsilon_{\mu\nu\lambda}
\partial_\nu A_\lambda) \, \delta (\partial \cdot {\tilde F}) $$
\be
=\, \int \,{\cal D}\ft \, {\cal D}\lambda_\mu \, {\cal D}\theta \,
\exp \left\{ i \int d^3 x [ \lambda_\mu ( \ft - \epsilon_{\mu\nu\rho}
\partial_\nu A_\rho )\,+ \, \theta \partial_\mu \ft ] \right\}
\label{expone}
\ee
where $\lambda_\mu$ and $\theta$ are Lagrange multipliers.
Integrating out $\ft$ in (\ref{expone}) yields
\begin{eqnarray}
& & \int \,{\cal D} \ft \, \delta (\ft - \epsilon_{\mu\nu\lambda}
\partial_\nu A_\lambda) \, \delta (\partial \cdot {\tilde F}) 
= \nonumber\\
& &  \int \, {\cal D}\lambda \, {\cal D}\theta \,
\delta (\lambda_\mu - \partial_\mu \theta) \exp \left( -i \int d^3 x \,
\lambda_\mu \epsilon_{\mu\nu\rho} \partial_\nu A_\rho \right) =
\nonumber\\
& & 
 \int {\cal D}\theta \, \exp \left( -i \int d^3 x \partial_\mu
\theta \, \epsilon_{\mu\nu\rho} \partial_\nu A_\rho \right)\;=\;
\int {\cal D}\theta \, \exp \left( i \int d^3 x \theta \,
\epsilon_{\mu\nu\rho} \partial_\mu \partial_\nu A_\rho \right) 
\nonumber\\
\label{exinone}
\end{eqnarray}
where the vanishing of $\fmn$ at infinity was used on the last
line, in order to ignore the surface contribution. We conclude,
after integrating out $\theta$ in $(\ref{exinone})$ that
$$\int \,{\cal D} \ft \, \delta (\ft - \epsilon_{\mu\nu\lambda}
\partial_\nu A_\lambda) \, \delta (\partial \cdot {\tilde F}) $$
\be
=\, \delta (\epsilon_{\mu\nu\rho} \partial_\mu \partial_\nu A_\rho )\;.
\label{pipo}
\ee
Thus the `1' behaves as a constant factor when inserted into
a functional integration over $A_\mu$ fields whose second
partial derivatives commute.

After insertion of the `1', the generating functional becomes
\be
{\cal Z}[J_\mu] \;=\; \int \,{\cal D} A_\mu \, {\cal D}\ft \,
\delta (\partial \cdot A)\, \delta (\partial \cdot {\tilde F})\,
\delta (\ft - \epsilon_{\mu\nu\lambda}\partial_\nu A_\lambda)
e^{- I (\ft) + \int d^3 x J_\mu (x) A_\mu (x) } \;.
\label{gfxz}
\ee
Now we observe that, introducing  the two delta-functionals
$\delta (\partial \cdot A)$ and
$\delta (\ft - \epsilon_{\mu\nu\lambda}\partial_\nu A_\lambda)$,
$A_\mu$ can be written in terms of $\ft$:
\be
A_\mu \;=\; -\epsilon_{\mu\nu\lambda} \frac{1}{\partial^2}
\partial_\nu {\tilde F}_\lambda \;,
\label{fta}
\ee
and the dependence on $A_\mu$ (only from the source term) can be
completely erased by replacing it by its expression (\ref{fta})
in terms of $\ft$. The $A_\mu$ field is thus integrated out, yielding
for ${\cal Z}$ the expression:
\be
{\cal Z}[J_\mu] \;=\; \int \,{\cal D}\ft \, \delta (\partial
\cdot {\tilde F})\,
\exp
\left({- I (\ft) - \int d^3 x J_\mu \epsilon_{\mu\nu\lambda}
\partial_\nu \partial^{-2} {\tilde F}_\lambda } 
\right)\;,
\label{zfin}
\ee
which contains only $\ft$ as a dynamicaly variable, and may be
thought of as the generating functional for a theory describing
the dynamics of a pseudovector field $\ft$, with  the
constraint $\partial \cdot {\tilde F}=0$. We note that,
because of the form of the source term in (\ref{zfin}),
there is a simple relation between Green's functions for
$\ft$ and the ones for $A_\mu$:

\begin{eqnarray}  \langle A_{\mu_1}(x_1) A_{\mu_2}(x_2)
\!\!\!\! & \cdots&\!\!\!\!
A_{\mu_n}(x_n)\rangle = \epsilon_{\mu_1 \nu_1
\lambda_1}\partial^{x_1}_{\nu_1}\partial_{x_1}^{-2}
\;
\epsilon_{\mu_2 \nu_2
\lambda_2}\partial^{x_2}_{\nu_2}\partial_{x_2}^{-2}
\cdots
\nonumber\\
& & \cdots
\epsilon_{\mu_n \nu_n
\lambda_n}\partial^{x_n}_{\nu_n}\partial_{x_n}^{-2}
\langle
F_{\lambda_1}(x_1) F_{\lambda_2}(x_2) \cdots F_{\lambda_n}(x_n)
\rangle \;.
\label{relgf}
\end{eqnarray}

Although a naive look at (\ref{zfin}) may suggest that it is
tantamount to a gauge fixed version for some gauge-invariant
theory, this is not necessarily the case, as the form
of the `action' $I$ for the pseudovector field is not
`gauge invariant' in this sense.

In order to do actual calculations with the theory defined
in terms of $\ft$, a  set of Feynman rules should be
defined.
It is convenient to introduce a Lagrange multiplier
field $\theta$ in order to deal with the delta-functional
$\partial \cdot {\tilde F}$,
and also to add a source term for $\theta$, since $\ft$
and $\theta$ are coupled. We add a source term for $\ft$ (not to be
confused with the source for $A_\mu$),
since the Green's functions for $A$ may be obtained by
applying (\ref{relgf}) to the $\ft$'s Green's functions.

Thus the generating functional we define is
\be
{\cal Z} \;=\; \int \,{\cal D}\ft \, {\cal D}\theta \,
\exp \left\{ - I(\ft) + \int d^3 x [ i \theta
\partial \cdot {\tilde F} \,+\,J_\mu \ft + j_\theta
\theta ] \right\}
\label{ztot}
\ee
and Euclidean correlation functions are simply obtained
by functional differentiation.
Free propagators are obtained from evaluation of the
Gaussian integral corresponding to a quadratic action,
\be
I(\ft) \; \equiv \; I_0(\ft) \; = \; \int d^3 x
\oh \; \ft \, \ft \;.
\label{ifree}
\ee
It is immediate to extract  the free propagators that follow
from (\ref{ztot}) with the action (\ref{ifree})
$$\langle {\tilde F}_\mu {\tilde F}_\nu \rangle \;=\;
(\delta_{\mu\nu} - \frac{k_\mu k_\nu}{k^2})$$
$$\langle \theta \theta \rangle \;=\; \frac{1}{k^2}$$
\be
\langle \ft \theta \rangle \;=\; \frac{k_\mu}{k^2}
\ee

The field strenght formulation in the Abelian case can be applied
to the study of the interaction between magnetic monopoles~\cite{hal}.
This can be achieved by replacing in
(\ref{zfin}) the delta functional of the Bianchi identity
by a delta functional that enforces $\partial_\mu \ft$ to equal a
monopole density $\rho$:
\be
{\cal Z}[J_\mu] \;=\; \int \,{\cal D}\ft \, \delta (\partial
\cdot {\tilde F} - \rho (x))\,
e^{- I (\ft) - \int d^3 x J_\mu \epsilon_{\mu\nu\lambda}
\partial_\nu \partial^{-2} {\tilde F}_\lambda } \;.
\label{zmon}
\ee
An argument similar to the one used at the beginning of this section
shows that the $`1'$ in this case reduces to a constant factor when put
inside a functional integral where the partial derivatives
of $A_\mu$ don't commute, but rather satisfy

\begin{equation}
\epsilon_{\mu\nu\lambda}\partial_\mu \partial_\nu A_\lambda = \rho.
\label{mon}
\end{equation}
It should be stressed at this point that actual monopoles are
in fact   static configurations in Minkowski $3+1$ dimensional
space-time while
the  present discussion corresponds to $3$-dimensional
 Euclidean
space-time. Then, it should be more appropriate to consider
these configurations as instantons(i.e.  points in space time) rather than
as  particles (monopoles). The fact that the magnetic field resulting
from  (\ref{mon}) has the typical monopole behavior  justifies
however
our terminology.

The interaction action for the monopoles is then deduced by
setting $J_\mu = 0$, and integrating out over $\ft$. This
yiels an effective description for the interaction between
monopoles, in terms of the scalar field $\theta$ and the
density $\rho$. Of course this will represent the interaction
{\em due to the gauge field\/}, as we are not giving rho any
dynamics. It could, of course, be introduced at the end, and
this will not affect the following discussion.

Of course, in the Born-Infeld case the integral over $\ft$
is not Gaussian, but, as there are no derivatives of $\ft$, we
see that the corresponding action is of the `ultralocal'
kind, since the integral  decomposes into an infinite
product of uncoupled integrals (one for each space-time point).
The result of performing the functional integral will then be
an infinite product (one factor for each spacetime point), of
normal, three dimensional integrals. Denoting by ${\cal Z}(\rho)$
the functional integral for $J=0$, we have
\begin{eqnarray}
{\cal Z}(\rho) &=& \int {\cal D} \theta
\exp (i \int d^3x \theta \rho)
\Pi_x
\left\{ \int d \ft \,
\exp[ - \beta^2 ( R(\frac{\ft^2}{\beta^2}) - 1 ) + \right. \nonumber\\
& &  \left.i \ft \partial_\mu
\theta (x) ] \right\}
\label{cortar}
\end{eqnarray}
where $\ft$ is understood as $x$ independent, and the product runs
over all the spacetime points.  To obtain the integral over
$\ft$ we just need to know the $x$-dependent, normal (rather than
functional)
integral
\be
\exp [ - W(\theta (x)) ] \,\equiv\,\int d \ft \,\exp[-\beta^2 (
R(\frac{\ft^2}{\beta^2})
- 1 )+ i \ft \partial_\mu \theta (x) ] \;.
\label{difi}
\ee
The range of integration over $\ft$ should be restricted to be
inside a sphere of radius $\beta$, if that is the value of the
absolute field. We assume that kind of cutoff, if present, has
been included into the function $R$.
Then,
\be
{\cal Z}(\rho)\;=\;\int {\cal D} \theta
\exp (i \int d^3x \theta \rho)
\Pi_x  \exp [ - W(\theta (x)) ] \;.
\ee

The integral (\ref{difi}) can of course be rewritten in spherical
coordinates:
\begin{eqnarray}
e^{- W(\theta (x))} & = &
2 \pi \int_0^\infty d |\ft| |\ft|^2 \,
\int_0^\pi d \omega \sin \omega \exp[ - \beta^2 (
R(\frac{\ft^2}{\beta^2}) - 1 )
+ \nonumber \\
& & i |\ft| |\partial_\mu \theta (x)| \cos \omega ],
\label{corto}
\end{eqnarray}
where $|\ft| = \sqrt{{\tilde F}^2}$ and $|\partial_\mu \theta (x)| =
\sqrt{(\partial\cdot\theta)^2}$. Performing the angular
integration yields
\be
e^{- W(\theta (x))} = 4 \pi \int_0^\infty d |\ft| |\ft| \,
\frac{\sin [  \sqrt{{\tilde F}^2}
\sqrt{(\partial\cdot\theta)^2}]}{\sqrt{
(\partial\cdot\theta)^2}}
\exp[ - \beta^2 ( R(\frac{\ft^2}{\beta^2}) - 1 )]
\label{intm}
\ee

For the sake of comparison, we evaluate first the integral (\ref{intm}) in
the Maxwell case ($R(x)=x$), what yields for the ${\cal Z}(\rho)$ the
expression
\be
{\cal Z}(\rho)\;=\;\int {\cal D} \theta
\exp (i \int d^3x \theta \rho)
\exp \left\{ - \frac{1}{4} \int d^3 x \partial_\mu \theta 
\partial_\mu \theta  
\right\} \;.
\ee
This shows that, for the Maxwell case, the monopoles shall interact 
through $1/r$ propagators, the three dimensional equivalent of the
usual logarithmic interaction between vortices for the $O(2)$ model on 
the plane. 
Unfortunately, the integral (\ref{intm}) cannot be evaluated exactly
for the Born-Infeld case. However, to understand at least qualitatively the
kind of modification introduced by that kind of action on the
interaction between monopoles, we shall use a function $R$ which allows
us to perform the integral exactly, 
\be
R (x)\;=\; \sqrt{x} \;.
\ee
Of course, this function  differs from the one
corresponding to the Born Infeld theory but 
it has in common with it that, for large fields,
the action grows linearly in contrast with what happens
in the Maxwell case. The result for ${\cal Z}_\rho$ is,
\be
{\cal Z}(\rho)\;=\;\int {\cal D} \theta
\exp (i \int d^3x \theta \rho)
\exp \left\{ - \int d^3 x \beta^2 \ln [ \partial_\mu \theta 
\partial_\mu \theta  + \beta^2]^2
\right\} \;
\ee
so that one can see 
that the interaction between monopoles is strongly different to the
one induced for the usual, Maxwell case.  

~

\underline{Acknowledgements}: F.A.S. is
partially  suported by CICBA and CONICET,
Argentina and a Commission of the European Communities
contract No:C11*-CT93-0315. C.D.F. is supported by CONICET.

\newpage

\vspace{3 cm}

\epsfxsize=6.in
\epsffile[100 100 600 700]{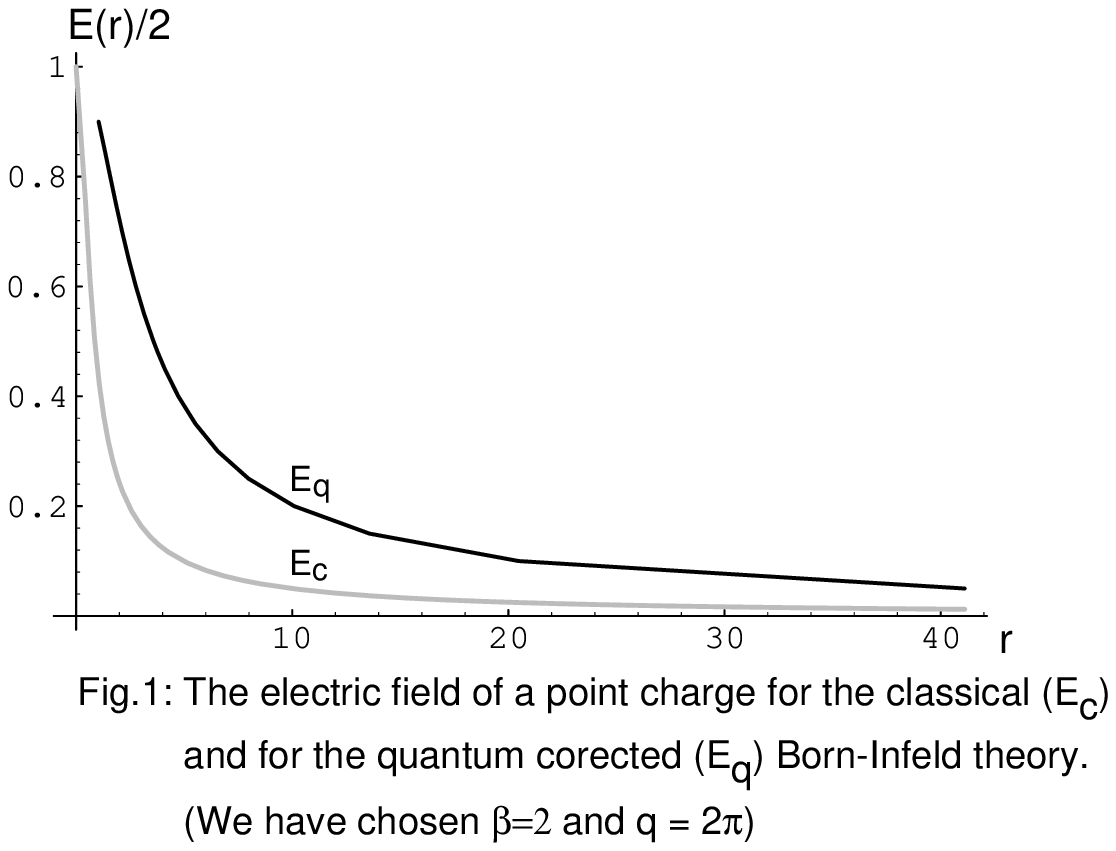}

\end{document}